\documentclass[aps,prb]{revtex4}
\usepackage{amsmath}
\usepackage{graphicx}

\begin{document}

\title{ Enhancement of   
 non-contact friction  between metal surfaces induced by the electrical double layer }

\author{A.I. Volokitin$^{*}$}

\affiliation{
Samara State Technical University,  443100 Samara, Russia}

\begin{abstract}

Casimir and electrostatic non-contact friction   between  two gold plates,  and a gold tip and a gold plate,   are calculated   taking into account the contribution of the  electrical double layer. It is shown that in an extreme-near field ($d<10$nm) the contribution from the electrical double layer leads to the enhancement of   non-contact  friction  by many orders of magnitude in comparison to the result of the conventional theory without    this contribution.    Casimir and electrostatic friction  dominate for short and large separations, respectively. The calculated electrostatic friction is in good agreement with  experimental data.  The results obtained open the way to detect the Casimir friction using Atomic Force Microscope.
\end{abstract}
\maketitle


\section{Introduction}

\vskip 5mm

\begin{figure}
\hskip 1.0cm
\includegraphics[width=0.8\textwidth]{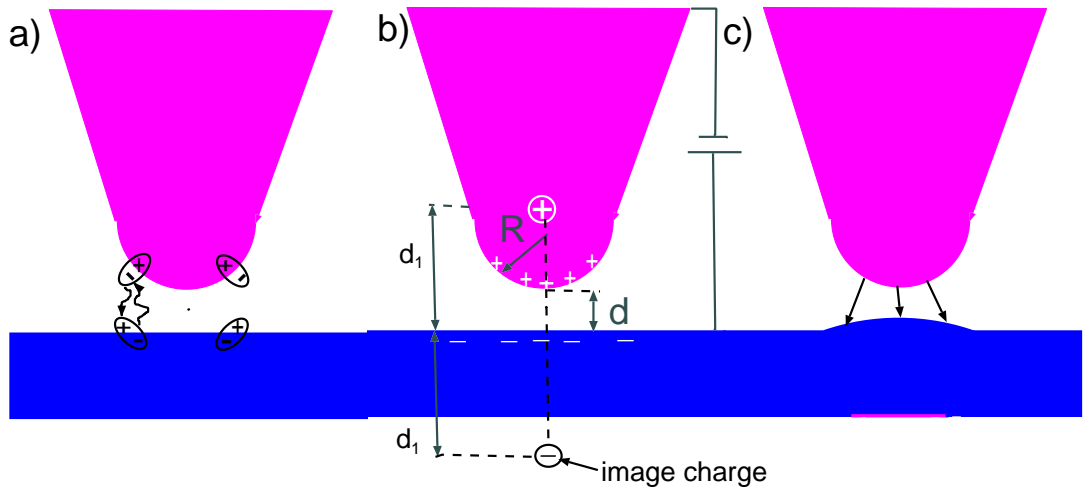}
%\vskip -5cm
\caption{ Origin of non-contact friction}
\label{SchemeNc_Fr}
\end{figure}

\begin{figure}
\hskip 5.0cm
\includegraphics[width=0.7\textwidth]{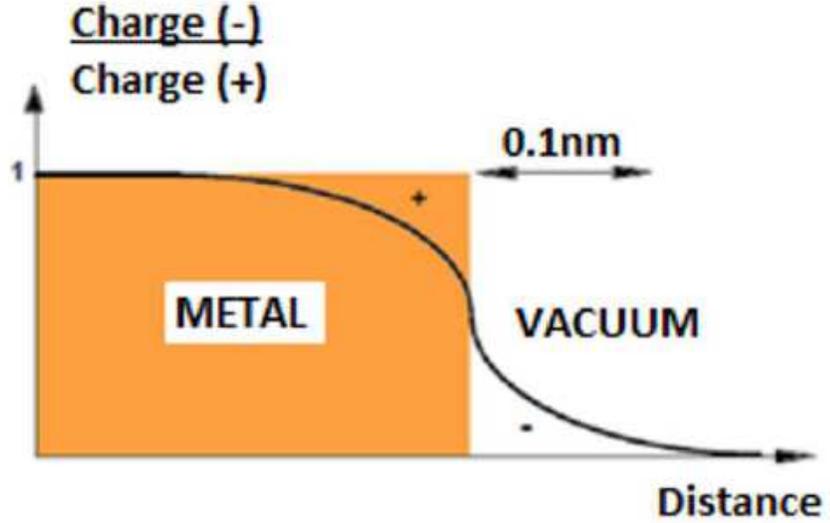}
%\vskip -5cm
\caption{ The electrical double layer  on metal surface (schematic)\cite{Kang2014ElectrochimActa}}
\label{SchemeDL}
\end{figure}

One of the important achievement   in  nanotribology is the discovery of non-contact friction between bodies without direct contact\cite{Stipe2001PRL,Kiesel2011NatMat,Kiesel2013NatMat,Volokitin2007RMP,Kiesel2015Nanoscience,Volokitin2017Book}. 
Beside of the fundamental significance of this phenomenon, it has practical application in biology and quantum computing for ultra sensitive force registration due to the link between friction and force fluctuations established by the fluctuation-dissipation theorem. Origin of non-contact friction is shown in Fig.\ref{SchemeNc_Fr}: (a) \textit{Casimir friction}. A fluctuating electromagnetic field created by quantum and thermal fluctuations of the current density in one body induces the current density in other body. The interaction between the fluctuating and induced current densities is an origin of the Casimir force. When the bodies are in relative motion the induced current lag behind and pull the fluctuating current back producing  the Casimir friction.   A theory of the  Casimir friction for relative sliding of two plates was developed in  Refs.\cite{Volokitin1999JPCM,VolokitinPRB2008} using  the fluctuational electrodynamics developed  by Rytov \textit{at.al} \cite{Rytov1953,Rytov1967,Rytov1987}.  The Casimir friction exists even at zero temperature due to  quantum fluctuations and is denoted as quantum friction\cite{Pendry1997JPCM}.  In contrast to the Casimir forces and radiative heat transfer, which were measured  in many experiments, the detection of the Casimir friction is still a challenging problem for experimentalists. However, the Casimir frictional drag force between quantum wells
and graphene sheets, and the current-voltage dependence
of nonsuspended graphene on the surface of the
polar dielectric SiO$_2$, were accurately described using the
theory of the Casimir friction\cite{Volokitin2001JPCM,Volokitin2013EPL,Volokitin2017Z_Natur,Volokitin2011PRL}. In Ref.\cite{Volokitin2016JETPLett,Volokitin2016PRB} it was shown that the Casimir frictional drag force produced by current density in a graphene sheet  can be detected  using an AFM. However, at present there is no direct mechanical detection of the Casimir friction. 
(b)  Interaction of a charged tip with an image charge produces an \textit{electrostatic friction} during motion of the tip. Theory of the electrostatic friction was developed in Refs.\cite{Volokitin2005PRL,Volokitin2006PRB}. Although the Casimir and electrostatic friction are both of electromagnetic  origin, the detailed mechanism is not totally clear.  In the conventional theory, where the reflection amplitudes are determined by the Fresnel's formulas, calculated  friction is  many orders of magnitudes smaller than experimentally observed values\cite{Stipe2001PRL,Kiesel2011NatMat,Kiesel2013NatMat}. However the friction is enhanced by many orders of magnitudes by ionic adsorbates\cite{Volokitin2005PRL,Volokitin2006PRB,Volokitin2003PRL}.(c) The van der Waals and electrostatic interactions between fluctuating surface displacements produce a time-dependent stresses acting on the surfaces that excite the acoustic phonons and produce \textit{phonon} heat transfer and  friction for moving bodies.

In this article the Casimir and electrostatic friction are calculated between gold plates, and a gold tip and a gold plate taking into account  of surface charge density related with the electrical double layer.  Metal consists of   positive ions occupying the positions of the crystal lattice and mobile electrons. The electrons can not  move away  from the crystal because the positive nucleus exerts an attractive force. Thus,  the surface of a clean metallic material can be considered as the superposition of two thin layers; one positively charged below the surface of the solid and the other negatively charged adjacent on the surface. This charge separation zone constitutes the electrical double layer  (Fig.\ref{SchemeDL}). Approximating the double layer by two opposite charged planes, the plane charge density can be estimated from the relation $4\pi\sigma_sed_0=\Delta\varphi$ where for gold the separation between planes $d_0\approx 0.1$nm and  the potential step due to the double layer\cite{Volokitin2021PRB} $\Delta\varphi\approx 4.3$eV, thus $\sigma_s\approx 0.38$Cm$^{-2}$. It is shown that in an extreme near-field ($d<10$nm) the Casimir and electrostatic friction are enhanced by many orders of the magnitudes in comparison to the case when the electrical double layer  is not taken into account, and they can be measured using the present state of art equipment. Giant enhancement of heat transfer between gold surfaces in an extreme near-field due to the electrical double layer was studied   in Ref.\cite{Volokitin2021PRB}.

\section{Casimir friction}

According to the theory of the Casimir friction\cite{Volokitin1999JPCM,VolokitinPRB2008,Volokitin2007RMP,Volokitin2017Book}, 
for two plates separated  by a vacuum gap  with thickness $d$ and sliding with relative velocity $v$,  for  $d<\lambda_T=c\hbar/k_BT$ and non relativistic velocity $v\ll c$, the frictional force  between surfaces   is dominated by the contribution from evanescent electromagnetic waves and determined by 

\begin{equation}
f_x =\hbar \int_0^\infty \frac{d\omega}{2\pi}\int_{q>k}\frac{d^2qq_x}{(2\pi)^2}e^{-2k_zd}
\left[\frac{4
\mathrm{Im}R_{1p}(\omega,q)\mathrm{Im}R_{2p}(\omega^{\prime},q) }{\mid 1-e^{-2
k_z d}R_{1p}(\omega,q)R_{2p}(\omega^{\prime},q)\mid ^2}\left[n_2(\omega_2^{\prime})-n_1(\omega)\right]+(p\rightarrow s)\right],
\label{Heat}
\end{equation}
where $n_i(\omega)=[e^{\hbar\omega/k_BT_i}-1]^{-1},$
 $\omega^{\prime}=\omega-q_xv$, $R_p$ and $R_s$ are the reflection amplitudes for $p-$ and $s-$ polarized electromagnetic waves, $k=\omega/c$, $k_z=\sqrt{q^2-k^2}$, $q>\omega/c$ is the component of the wave vector parallel to the surface. To linear order in the sliding velocity $f_x=\gamma v$ where the friction coefficient at $T_1=T_2=T$ is determined by \cite{Volokitin1999JPCM,VolokitinPRB2008, Volokitin2007RMP,Volokitin2017Book}
\begin{equation}
\gamma_{rad}=\frac{\hbar^2}{8\pi^2k_BT} \int_0^\infty \frac{d\omega}{\mathrm{sinh}^2(\hbar\omega/2k_BT)}\int_{0}^{\infty}dk_zk_z(k_z^2+k^2)e^{-2k_zd}
\left[\frac{
\mathrm{Im}R_{1p}(\omega,q)\mathrm{Im}R_{2p}(\omega,q) }{\mid 1-e^{-2
k_z d}R_{1p}(\omega,q)R_{2p}(\omega,q)\mid ^2}+(p\rightarrow s)\right].
\label{frcrad}
\end{equation}
In the presence of the surface charge density, that can be due to a potential difference or  the electrical double layer,  the reflection amplitude
 for the  $\textit{p}$ -polarized   electromagnetic waves is determined by  \cite{Volokitin2019JETPLett}
\begin{equation}
R_p=\frac{i\varepsilon k_z  -k_z^{\prime} +4\pi iq^2\alpha_{s}\varepsilon}{i\varepsilon k_z  +k_z^{\prime} -4\pi iq^2\alpha_{s}\varepsilon}
\label{rcp}
\end{equation}
where $k_z^{\prime}=\sqrt{\varepsilon k^2-q^2}$, $\varepsilon$ is the dielectric function of the substrate, and $\alpha_s$ is the normal component of the surface dipole susceptibility. Without taking into account the contribution of the electric double layer, Eq.(\ref{rcp})  reduces to the Fresnel's formula for the reflection amplitude for $p$-polarized electromagnetic waves
\begin{equation}
R_p=\frac{i\varepsilon k_z  -k_z^{\prime}}{i\varepsilon k_z  +k_z^{\prime} }.
\label{frcp}
\end{equation}
Due to the screening by electrons in surface layer the interaction of an external electromagnetic field with lower   of the double layer plane can be neglected. In this case   $\alpha_s=\sigma_s^2M$ where  
$M$ is the surface mechanical susceptibility that determines the surface displacement under the action of external mechanical stress: $u=M\sigma_{zz}^{ext}$. In the  elastic continuum model
\cite{Persson2001JPCM}
\begin{equation}
M=\frac{i}{\rho c_t^2}\left(\frac{\omega}{c_t}\right)^2\frac{p_l(q,\omega)}{S(q,\omega)},
\end{equation}
where
\[
S(q,\omega)=\left[\left(\frac{\omega}{c_t}\right)^2-2q^2\right]^2+4q^2p_tp_l,
\]
\[
p_t=\left[\left(\frac{\omega}{c_t}\right)^2-q^2+i0\right]^{1/2}, \,\,p_l=\left[\left(\frac{\omega}{c_l}\right)^2-q^2+i0\right]^{1/2},
\]
where  $\rho$,  $c_l$, and
$c_t$ are the mass density of the medium, the velocity of the longitudinal and transverse acoustic waves. A surface charge density and dipole moment is  also induced by  the applied potential difference\cite{Volokitin2019JETPLett}. Due to the screening of the parallel component of the electric field the contribution of the electrical double layer to the reflection amplitude for $s$-polarized electromagnetic waves can be neglected and for these waves the reflection amplitude is determined by the the Fresnel's formula
\begin{equation}
R_s=\frac{i k_z  -k_z^{\prime}}{i k_z  +k_z^{\prime} }.
\label{frcs}
\end{equation}

\section{Phononic friction}
 
The electrostatic and van der Waals interactions between surfaces 1 and 2  results in  stresses that act on surfaces 1 and 2.  If    $u_1$ and  $u_2$ denote the surface displacements of solid 1 and 2 then   
equations\cite{Volokitin2019JETPLett,Volokitin2020JPCM} 
\begin{equation}
\sigma_1(\omega)=au_{1}(\omega)-bu_{2}(\omega^{\prime}),
\label{sigma1}
\end{equation}
\begin{equation}
\sigma_2^{\prime})=au_{2}(\omega^{\prime})-bu_{1}(\omega),
\label{sigma2}
\end{equation}
where $\omega^{\prime}=\omega-q_xv$,  
\[
a=\frac{H}{2\pi d^4} + q\sigma_s^2\mathrm{coth}qd,\,b=\frac{H}{4\pi}\frac{q^2K_2(qd)}{d^2} + \frac{q\sigma_s^2}{\mathrm{sinh}qd},
\]
where the first and second terms are due to the van der Waals and 
electrostatic interactions, respectively, $H$ is the Hamaker constant, 
$K_2(x)$ is the modified Bessel function of the second kind and second order. 
The   surface displacements  due to thermal and quantum fluctuations are determined by \cite{Volokitin2019JETPLett,Persson2011JPCM}
\begin{equation}
u_1(\omega)= u_1^f(\omega)+M_1(\omega)[au_1(\omega)-bu_2(\omega^{\prime})],
\label{eqg}
\end{equation}
\begin{equation}
u_2(\omega^{\prime})= u_2^f(\omega^{\prime})+M_2(\omega^{\prime})[au_2(\omega^{\prime})-bu_1(\omega)],
\label{eqd}
\end{equation}
where according to the fluctuation-dissipation theorem, the spectral density of fluctuations of the surface displacements is determined by  \cite{LandauStatisticalPhysics}
\begin{equation}
\langle|u_i^f|^2\rangle = \hbar \mathrm{Im}M_i(\omega,q)\coth\frac{\hbar\omega}{2k_BT_i}
\label{fdt}
\end{equation}
where $M_i$ is the mechanical susceptibility for surface $i$.  The friction force can be calculated from equation\cite{Volokitin2017Book,Volokitin2002PRB}
\begin{equation}
f_xv=Q_1 + Q_2
\label{fv}
\end{equation}
where $Q_1$ and $Q_2$  are the heat generated in surfaces 1 and 2   which are determined by the rate of the work of the mechanical stress that act on  surfaces 1 and 2 in the rest reference  frame of surfaces 1 and 2
\begin{equation}
Q_1=
\int_{-\infty}^\infty\frac{d\omega}{2\pi}\int\frac{d^2q}{(2\pi)^2}\omega\mathrm{Im}
\langle u_1\sigma_1\rangle,
\label{Q1}
\end{equation}
\begin{equation}
Q_2=
\int_{-\infty}^\infty\frac{d\omega}{2\pi}\int\frac{d^2q}{(2\pi)^2}\omega^{\prime}\mathrm{Im}
\langle u_2\sigma_2\rangle
\label{Q2}
\end{equation}
From Eqs.(\ref{sigma1}-\ref{Q2}) we get 
  \begin{equation}
f_x=4\int_{0}^\infty\frac{d\omega}{2\pi}\int\frac{d^2q}{(2\pi)^2}q_x
\frac{b^2\mathrm{Im}M_1(\omega)\mathrm{Im}M_2(\omega^{\prime}) }{\mid (1-aM_1(\omega) )(1-aM_2(\omega^{\prime} )-b^2M_1(\omega)M_2(\omega^{\prime})\mid^2}\left[n_2(\omega^{\prime})-n_1(\omega)\right].
\label{frfel}
\end{equation}
 At $T_1=T_2=T$ the phonon friction coefficient is given by 
\begin{equation} 
\gamma_{ph}=\frac{\hbar^2}{8\pi^2k_BT} \int_0^\infty \frac{d\omega}{\mathrm{sinh}^2(\hbar\omega/2k_BT)}\int_{0}^{\infty}dqq^3\frac{b^2\mathrm{Im}M_1(\omega)\mathrm{Im}M_2(\omega) }{\mid (1-aM_1(\omega) )(1-aM_2(\omega)-b^2M_1(\omega)M_2(\omega)\mid^2}.
\label{frcph}
\end{equation}

\section{Numerical results}

\begin{figure}
\includegraphics[width=0.5\textwidth]{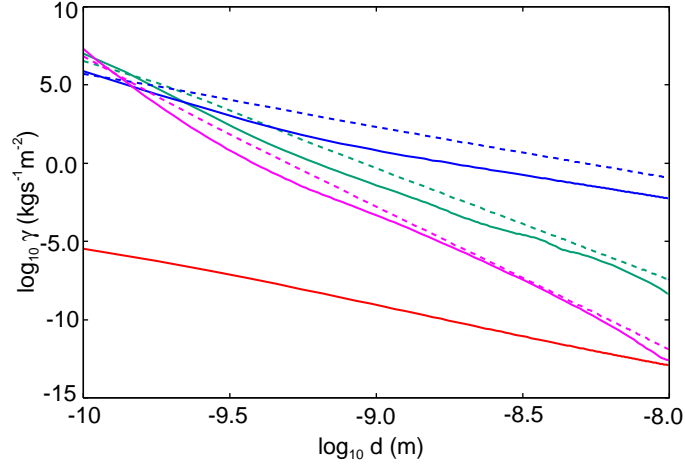}
\caption{The dependence of the  Casimir friction  coefficient   between two gold plates  on the distance between them  at $T=300$K. The blue   lines  is  the contributions from fluctuating dipole moment of surface electrical double layer at the double layer potential step $\Delta\varphi = 4.3$eV.   The brown lines is  the contributions  from the fluctuating dipole moment induced on surfaces by the potential difference $\varphi=10$V. The pink line is  the contribution from van der Waals interaction between  fluctuating surface displacements.   The Full and dashed lines are  the contributions from the bulk and surface (Rayleigh) phonon modes. The red line is the  result of the conversional theory of the Casimir friction without the contributions from the  fluctuations of the surface displacements. \label{Plate-Plate}}
\end{figure}

Fig.\ref{Plate-Plate} shows the dependence of friction coefficient for the Casimir friction   between two gold plates on the distance between them for different mechanisms at   $T=300$K. For gold  $c_l=3240$ms$^{-1}$, $c_t=1200$ms$^{-1}$, $\rho=1.9280\times10^4$kgm$^{-3}$, $H=34.7\times 10^{-20}$J (see Ref.\cite{Hamaker2015})  and dielectric function\cite{Chapuis2008PRB}
\begin{equation}
\varepsilon=1-\frac{\omega_p^2}{\omega^2+i\omega\nu},
\end{equation}
where  $\omega_p=1.71\times10^{16}$s$^{-1}$, $\nu=4.05\times10^{13}$s$^{-1}$ .      In Fig.\ref{Plate-Plate}  the blue  lines  are  for the Casimir friction   associated with the electrical double layer and  $p$-polarized waves. The green and pink  lines are for   the phonon friction associated with the van der Waals and the electrostatic interaction between the fluctuating surface displacements at the potential difference  $\varphi=10$V. The red line is the theory result  for the Casimir friction without  the contribution from the double layer.   Full and dashed lines are for the contributions from the bulk and surface (Rayleigh) acoustic modes. The mechanical  susceptibility has a pole at $\omega=\omega_s=c_sq=\xi c_tq$, where  $\omega_s$ and $c_s$ are the frequency and the propagation velocity of the Rayleigh surface waves,   where for gold   $\xi=0.94$. Near the pole at $\omega\approx \omega_s$ the mechanical susceptibility can be written as\cite{Volokitin2020JPCM}
\begin{equation}
M=-\frac {c}{\rho c_t(\omega-c_sq+i\gamma)},
\label{mpol}
\end{equation}
where
for gold $c=0.36$ and  $\gamma$ is the damping constant for surface wave for which can be used estimate\cite{Volokitin2020JPCM,Pendry2016PRB} $\gamma=0.17\omega$.  In the  extreme near-field ($d<10$nm) the double layer contribution leads to an enhancement of the Casimir  friction by many orders of the magnitude in comparison with the  theory which does not include this contribution.

Vibration    of a tip charged by bias  voltage $\varphi$ parallel to  metallic substrate produces an electrostatic friction.   
In the case when the tip  is a section of the cylindrical surface with the radius of curvature $R$ and with the width $w$  the friction coefficient is the same as for a wire located at $d_1= \sqrt{(d+R)^2-R^2}$ and with the charge per unit length $Q=C\varphi $, where $C^{-1}=2\ln
[(d+R+d_1)/R]$, and is given by \cite{Volokitin2005PRL,Volokitin2006PRB,Volokitin2017Book}
\begin{equation}
\Gamma_c =\lim_{\omega \to 0}2C^2\varphi^2w\int_0^\infty
dqqe^{-2qd_1}\frac{\mathrm{ Im}R_p(\omega ,q)}{\omega}
\label{biasone}
\end{equation}
where  $R_p$ is the reflection amplitude for $p$-polarized waves. Using in Eq.(\ref{biasone}) the Fresnel's formula for the reflection amplitude 
gives the friction coefficient 
\begin{equation} 
\Gamma_c
^{Fresnel}=\frac{w\nu \varphi^2}{16 \omega_p^2 d^2}.  \label{biasthree}
\end{equation}
For gold with $ w=7\cdot 10^{-6}$
m,  $d=20$ nm and $\varphi=1$V, Eq.(\ref{biasthree})
gives $\Gamma_c =1.69\cdot 10^{-20}$ kg/s that is eight orders of
magnitude smaller than the experimental value $3\cdot
10^{-12}$kg/s from Ref.\cite{Stipe2001PRL}. In the case when the surface double layer is taken into account, for  complete screening, when the interaction  of the external electric field with  the subsurface charge distribution  is neglected, and $1/\varepsilon\ll 4\pi q\sigma_s^2M\ll 1$ the reflection amplitude (\ref{rcp}) can be approximated  by
\begin{equation}
R_p^{DL}\approx 1+8\pi q\sigma_s^2M
\label{DLrcp}
\end{equation}  
and the friction coefficient is determined by  
\begin{equation}
\Gamma_c =\lim_{\omega \to 0}16\pi \sigma_s^2C^2\varphi^2w\int_0^\infty
dqq^2e^{-2qd_1}\frac{ \mathrm{ Im}M(\omega ,q)}{\omega}.
\label{biasdl}
\end{equation}
The contribution to friction from bulk acoustic waves, which is determined by the integration for $q\leq\omega/c_t$, vanishes in the limit $\omega \rightarrow 0$. Thus at low frequency  the electrostatic friction coefficient is determined by Rayleigh surface waves. Substitution of (\ref{mpol}) in (\ref{biasdl}) for $R\gg d$ gives
\begin{equation}
\Gamma_c^{DL} = 0.14\frac{\sigma_s^2\varphi^2wR^{1/2}}{d^{3/2}\rho c_s^2c_t}.
\label{gal}
\end{equation} 
The friction coefficient has the same distance dependence as it was observed experimentally\cite{Stipe2001PRL}. At $R=1\mu$m  and   with the same parameters as above $\Gamma_c^{DL}=1.9\cdot 10^{-12}$kgs$^{-1}$  is in good agreement    experimental value $3\cdot
10^{-12}$kg/s from Ref.\cite{Stipe2001PRL}. 

For a spherical tip  the friction coefficient \cite{Volokitin2005PRL,Volokitin2006PRB,Volokitin2017Book}
\begin{equation}
\Gamma _{s}=\lim_{\omega \to 0}\frac{C^2\varphi^2}{2}\int_0^\infty
dqq^2e^{-2qd_1}\frac{\mathrm{Im}R_p(\omega ,q)}\omega  \label{biaseight}
\end{equation}
where
\begin{equation}
d_1=\pm \sqrt{3Rd/2+\sqrt{(3Rd/2)^2+Rd^3+d^4}},\,\,C=\frac{d_1^2-d^2}{2d}\label{d1}
\end{equation}
Substituting  in (\ref{C}) the Fresnel's formula for the reflection amplitude  and using Eq.(\ref{DLrcp})  gives
\begin{equation}
\Gamma_{s}^{Fresnel}=\frac{\sqrt{3R}\nu \varphi^2}{4\omega_p^2d^{3/2}}
\label{Gs}
\end{equation}
and 
\begin{equation}
\Gamma_s^{DL}=0.14\frac{R\sigma_d^2\varphi^2}{d\rho c_tc_s^2}
\label{Gds}
\end{equation}

\begin{figure}
\includegraphics[width=0.5\textwidth]{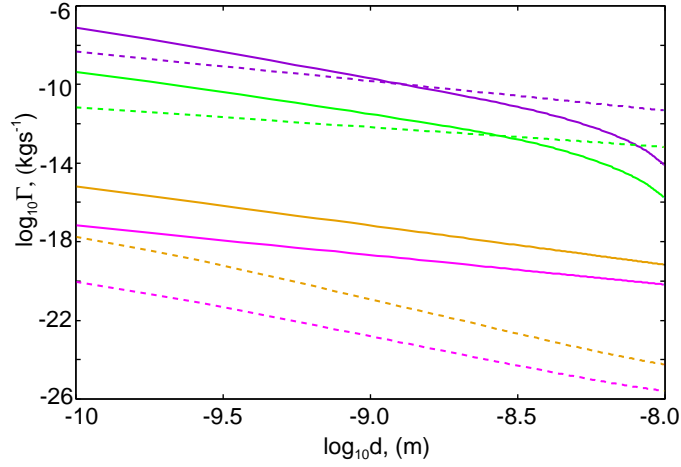}
\caption{The dependence of the Casimir (full lines) and electrostatic (dashed lines) friction coefficient   on the distance between a gold tip  and  a gold plate  at $T=300$K.  The blue and green lines include  the contributions from the electrical double layer for a spherical tip   with radius $R=1$$\mu$m and a cylindrical tip with radius $R=1$$\mu$m and width $w=7$$\mu$m, respectively.   The brown and pink lines are the results of the conventional theory for a cylindrical and spherical tip, respectively.   \label{Tip_Plate}}
\end{figure}

Fig. \ref{Tip_Plate}  shows  the dependence of the friction coefficient associated with the  friction between a gold  tip    and a gold plate on the separation $d$. The friction coefficient  for the Casimir friction in the tip-plate configuration can be obtained from the plate-plate configuration using the proximity  (or Derjyaguin) approximation \cite{Parsegian2006}.  Within this approximation the friction coefficient  for a spherical tip with a radius $R$
\begin{equation}
\Gamma_{s}=2 \pi\int_0^Rd\rho\rho \gamma(z(\rho))
\label{tipplane}
\end{equation}
and for a cylindrical tip ith a radius $R$ and a width $w$
\begin{equation}
\Gamma_{c}=2 w\int_0^Rd\rho \gamma(z(\rho))
\label{tipplane}
\end{equation} 
where  $\gamma(z)$ is the friction coefficient  between two plates separated by a distance $z(\rho)=d+R-\sqrt{R^2-\rho^2}$ denoting the tip-surface distance as a function of the distance $\rho$ from the tip symmetry. In the extreme-near field the friction coefficient $\gamma \sim 10^{-12}-10^{-13}$kgs$^{-1}$ which can be measured with the present state-of-the art equipment.
The blue and green lines include the contribution from the electrical double layer, while the brown and pink lines without this contribution, for a cylindrical and spherical tip, respectively. The cylindrical tip has the radius $R=1\mu$m and the width $w=7\mu$m, the spherical tip has the radius $R=30$nm.  For $d<10$nm the contribution from the electrical double layer produces  the friction coefficient $\gamma \sim 10^{-10}-10^{-12}$kgs$^{-1}$ that is  by many orders of the magnitude larger than the friction coefficient without this   contribution.  This friction  can be measured with the present state-of-the art equipment.

\section{Conclusion}
We have shown that the interaction of the electromagnetic field with the surface charge density of the electric double layer produces an enhancement  of the Casimir and electrostatic non-contact friction between a gold tip and a gold plate by many orders of magnitude in comparison with the   theory without taking into account of this contribution. For electrostatic friction, which are dominated by the excitation of Rayleigh surface waves, the theory is in good agreement with experimental data. The theory also shows that the electrical double layer contribution to the Casimir friction can be measured using AFM.

\vskip 0.5cm

The reported study was funded by RFBR according to the research project N\textsuperscript{\underline{o}} 19-02-00453

\vskip 0.5cm

$^*$alevolokitin@yandex.ru


\begin{thebibliography}{999}

\bibitem{Stipe2001PRL}   B. C. Stipe, H. J. Mamin, T. D. Stowe, T. W. Kenny, and D. Rugar, Noncontact Friction and Force Fluctuations between Closely Spaced Bodies, Phys. Rev. Lett. \textbf{87}, 096801(2001). 

\bibitem{Kiesel2011NatMat} M. Kisiel, E. Gnecco, U. Gysin, L. Marot, S. Rast and E. Meyer, Suppression of electronic friction on Nb films in
the superconducting state, Nature Materials \textbf{10}, 119(2011). 

\bibitem{Kiesel2013NatMat} M. Langer, M. Kisiel, R. Pawlak, F. Pellegrini, G. E. Santoro, R. Buzio, A. Gerbi, G. Balakrishnan, A. Baratoff, E. Tosatti
and Ernst Meyer, Giant frictional dissipation peaks and
charge-density-wave slips at the NbSe2 surface, Nature Materials \textbf{13}, 173 (2013).


\bibitem{Volokitin2007RMP}  A.I. Volokitin  and  B.N.J. Persson, Near field radiative heat transfer and noncontact friction, Rev. Mod. Phys.
\textbf{79}, 1291 (2007).


\bibitem{Kiesel2015Nanoscience} M. Kisiel, M. Samadashvili, U. Gysin, E. Meyer, Non-contact friction.   In: S. Morita, F. Giessibl, E. Meyer, R. Wiesendanger. (eds) Noncontact Atomic Force Microscopy. NanoScience and Technology. Springer, Cham.(2015)

\bibitem{Volokitin2017Book} A.I.Volokitin and B.N.J.Persson, Electromagnetic Fluctuations at the Nanoscale. Theory and Applications, (Springer, Heidelberg, 2017).

\bibitem{Volokitin1999JPCM} A. I. Volokitin and B. N. J. Persson. Theory of friction: the contribution from a fluctuating electromagnetic field. J. Phys.: Condens. Matter, \textbf{11}, 345 (1999).

\bibitem{VolokitinPRB2008} A. I. Volokitin and B. N. J. Persson. Theory of the interaction forces and the radiative heat transfer between moving bodies. Phys. Rev. B, \textbf{78}, 155437 (2008).

\bibitem{Rytov1953} S.M. Rytov, Theory of Electrical Fluctuations and Thermal Radiation (Akad. Nauk USSR, Moscow, 2953) [In Russian].

\bibitem{Rytov1967} M.I. Levin and S.M. Rytov, Theory of Equilibrium Thermal Fluctuations in Electrodynamics, (Nauka, Moscow, 1967) [in Russian].


\bibitem{Rytov1987} S.M. Rytov, Y.A.  Kravtsov,   and V.I. Tatarskii,  Principles of Statistical Radiophysics (Springer, Berlin, 1987). 

\bibitem{Pendry1997JPCM} J. B. Pendry. Shearing the vacuum - quantum friction. J. Phys.: Condens. Matter, \textbf{9}, 10301 (1997).

\bibitem{Volokitin2001JPCM} A.I.Volokitin and B.N.J.Persson, The frictional drag force between quantum wells mediated by a fluctuating electromagnetic field, J. Phys.: Condens. Matter \textbf{83}, 859(2001).

\bibitem{Volokitin2013EPL} A.I.Volokitin and B.N.J.Persson, Influence of electric current on the Casimir forces between graphene sheets, EPL, \textbf{103}, 24002 (2013).
\bibitem{Volokitin2017Z_Natur} A.I.Volokitin, Casimir Friction and Near-field Radiative Heat Transfer in Graphene Structures, Z. Naturforsch. A, \textbf{72}, 171(2017).

\bibitem{Volokitin2011PRL} A.I.Volokitin and B.N.J.Persson, Quantum friction, Phys. Rev. Lett., \textbf{106}, 094502 (2011).

 \bibitem{Volokitin2016PRB}A.I.Volokitin, Casimir frictional drag force between a SiO2 tip and a graphene-covered SiO2 substrate, Phys. Rev. B 94, 235450 (2016).



\bibitem{Volokitin2016JETPLett} A.I.Volokitin, Casimir Friction Force between a SiO2 Probe and a Graphene-Coated SiO2 Substrate, JETP Lett., \textbf{104}, 504 (2016).

\bibitem{Volokitin2005PRL} A.I. Volokitin and B.N.J. Persson, Adsorbate-Induced Enhancement of Electrostatic Noncontact Friction, 
Phys. Rev. Lett. \textbf{94}, 086104(2005).


\bibitem{Volokitin2006PRB} A. I. Volokitin, B. N. J. Persson, and H. Ueba, Enhancement of noncontact friction between closely spaced bodies by two-dimensional systems, Phys. Rev. B \textbf{73}, 165423(2006).



\bibitem{Volokitin2003PRL} A.I. Volokitin and B.N.J. Persson, Resonant Photon Tunneling Enhancement of the van der Waals Friction, Phys. Rev. Lett. \textbf{91}, 106101(2003). 


\bibitem{Kang2014ElectrochimActa}J. Kang, J. Wen, S. H. Jayaram, A. Yu, and X. Wang, Development of an equivalent circuit model for electrochemical double layer capacitors (EDLCs) with distinct electrolytes, Electrochim. Acta, \textbf{115}, 587 (2014).


\bibitem{Volokitin2021PRB} A. I. Volokitin, Electric double layer effect in an extreme near-field heat transfer between metal surfaces, Phys. Rev. B,  \textbf{103}, L041403(2021).

 
\bibitem{Volokitin2019JETPLett} A.I. Volokitin,  Effect of an Electric Field in the Heat Transfer between Metals
in the Extreme Near Field, JETP Lett.,\textbf{109}, 749(2019).

\bibitem{Persson2001JPCM} B.N.J. Persson, Theory of rubber friction and contact mechanics, J. Chem. Phys. \textbf{115}, 3840 (2001).


\bibitem{Volokitin2020JPCM} A.I.Volokitin, Contribution of the acoustic waves to near-field heat transfer, J. Phys.: Condens. Matter \textbf{32}, 215001(2020).

\bibitem{Persson2011JPCM} B.N.J. Persson, A.I.  Volokitin,  \& H. Ueba, H. Phononic heat transfer across an
interface: thermal boundary resistance. J. Phys. Condens. Matter \textbf{23}, 045009 (2011).

\bibitem{LandauStatisticalPhysics}  L.D. Landau and  E.M. Lifshitz, Statistical Physics ( Volume 5 of A Course of Theoretical Physics ) Pergamon Press, Oxford, 1980.

\bibitem{Volokitin2002PRB} A.I.Volokitin and B.N.J.Persson, Dissipative van der Waals interaction between a small particle and a metal surface, Phys. Rev. B, \textbf{65}, 115419(2002).


\bibitem{Hamaker2015} P. Pinchuk and K. Jiang, 
Size-dependent Hamaker constants for silver and gold nanoparticles, 
Proc. SPIE 9549, Physical Chemistry of    Interfaces and Nanomaterials XIV, 95491J (2015).

\bibitem{Chapuis2008PRB} P.-O. Chapuis, S. Volz, C. Henkel, K. Joulain, and J.-J. Greffet, Effects of spatial dispersion in near-field radiative heat transfer between two parallel metallic surfaces, Phys. Rev. \textbf{77}, 035431 (2008).

\bibitem{Pendry2016PRB} J.B. Pendry, K. Sasihithlu, R.V. Craste, Phonon-assisted heat transfer between vacuum-separated surfaces, Phys. Rev. B \textbf{94}, 075414 (2016).


\bibitem{Parsegian2006} V. A. Parsegian, Van der Waals Forces - A Handbook for Biologists, Chemists, Engineers, and Physicists (Cambridge
University Press, New York, 2006).










\end{thebibliography}
\end{document}